\documentclass[12pt]{iopart}

\usepackage{iopams}
\usepackage{braket}
\usepackage{amsfonts}
\usepackage{color}
\usepackage[dvipsnames]{xcolor}
\usepackage{amssymb}
\usepackage{mathrsfs}
\usepackage{graphicx}
\usepackage{lmodern}
\usepackage{lineno}
\usepackage{hyperref}
\usepackage[normalem]{ulem}

\newcommand{\sgn}{\textup{sgn}}
\newcommand{\spann}{\textup{span}}
\newcommand{\arcoth}{\textup{arcoth}}
\newcommand{\beq}{\begin{eqnarray}}
\newcommand{\eeq}{\end{eqnarray}}

\def\keywords#1{\vspace{10pt}
     \begin{indented}
     \item[]\rm Keywords: #1\par
     \end{indented}}

\begin{document}



\title{Tomography in Loop Quantum Cosmology}
\author{Jasel Berra--Montiel$^{1,2}$ and
Alberto Molgado$^{1,2}$}

\address{$^{1}$ Facultad de Ciencias, Universidad Aut\'onoma de San Luis 
Potos\'{\i} \\
Campus Pedregal, Av. Parque Chapultepec 1610, Col. Privadas del Pedregal, San
Luis Potos\'{\i}, SLP, 78217, Mexico}
\address{$^2$ Dual CP Institute of High Energy Physics, Mexico}

\eads{\mailto{\textcolor{blue}{jasel.berra@uaslp.mx}},\ 
\mailto{\textcolor{blue}{alberto.molgado@uaslp.mx}}\ 
}


\begin{abstract}
We analyze the tomographic representation for the Friedmann-Robertson-Walker (FRW) model within the Loop Quantum Cosmology framework.  We focus on the Wigner quasi-probability distributions associated with Gaussian and Schr\"odinger cat states, and then, by applying a Radon integral transform for those Wigner functions, we are able to obtain the symplectic tomograms which define measurable probability distributions that fully characterize the quantum model of our interest.  By appropriately introducing the quantum dispersion for a rotated and squeezed quadrature operator in terms of the position and momentum, we efficiently interpret the properties of such tomograms, being consequent with Heisenberg's uncertainty principle.  We also obtain, by means of the dual tomographic symbols, the expectation value for the volume operator, which coincides with the values reported in the literature.  We expect that our findings result interesting as the introduced tomographic representation may be further benefited from the well-developed measure techniques in the areas of Quantum optics and  Quantum information theory.
\end{abstract}



\section{Introduction}

The tomographic representation stands for a probabilistic interpretation of quantum mechanics where quantum states are defined through integral transforms of the density operators intending to obtain precise measures of the matrix elements belonging to those states~\cite{Manko}, \cite{Tombesi}, \cite{Simoni}, \cite{Manko4}, \cite{Asorey}.  In particular, here we are
interested in the tomographic representation obtained by means of the Radon integral transform of the Wigner quasi-probability function (or simply Wigner function) associated with the density operator within the context of  Phase space Quantum Mechanics \cite{Phasespace}, \cite{Bayen}, \cite{Bordemann}.  The obtained tomographic probability distribution, also called marginal distribution or simply a tomogram, results in a positive probability distribution, in contrast to the Wigner function within the phase space representation.  As expected, the standard tomographic representation results equivalent to either the Schr\"odinger or the Heisenberg representations being, however, a notorious distinction the generalized Fokker-Planck equation that defines quantum evolution~\cite{Manko3}. The tomographic representation has been extensively used, in particular in the fields of quantum optics and quantum information theory, to describe quantum states employing the tomograms associated with a set of well-defined observables,  allowing a natural discussion of relevant issues such as quantumness, superposition, entanglement,  quantum state reduction, and the reconstruction of coherent states, among others~\cite{Manko4}, \cite{Manko3}, \cite{Manko2}, \cite{DAriano}, \cite{Helsen}.  Tomographic techniques have been also recently introduced in the contexts of Quantum Field Theory~\cite{TQFT}, \cite{TQFT2}, Statistical Mechanics~\cite{Mendes} and the quantum cosmological scenario~\cite{Capozzielo}, \cite{Capozzielo2}, \cite{Stornaiolo}.

From our particular interest, we focus here on the introduction of the tomographic representation within the Loop Quantum Cosmology (LQC) scheme. The Loop Quantum Gravity (LQG) approach ~\cite{Improved}, \cite{MLQC}, \cite{Robustness}, \cite{LQCSR} (see also~\cite{Brief} for a state-of-the-art review), enforces diffeomorphism covariance at the quantum level by applying a non-regular representation of the canonical commutation relations which, for minisuperspace models, results in the so-called polymer representation for LQC. Recently, the polymer representation of quantum mechanics was analyzed from the phase space perspective, obtaining a Wigner quasi-probability function constructed by means of cylindrical functions defined on the Bohr compactification of the real line~\cite{Fewster}, \cite{Perlov}.  Further, in~\cite{Polymer}, \cite{Quasi} such Wigner function, together with a well-defined star-product, is recovered as a distributional limit of the Schr\"odinger representation  (see also \cite{ThiemannDQ1},~\cite{ThiemannDQ2} for  an analogous description from a different perspective). Nevertheless, as mentioned above, the Wigner function may acquire negative values on certain domains of the phase space.  In consequence, with the purpose to obtain an unambiguous probability distribution, we analyze the tomographic representation associated with the homogeneous and isotropic Friedmann-Robertson-Walker (FRW) quantum model emerging within the LQC framework. In order to determine the quantum tomograms, our strategy is to apply the Reduced Phase Space (RPS) approach of LQC \cite{EnergyBB}, \cite{TurningBB}, \cite{Algebraic}, where the physical phase space is identified by solving, at the classical level, the Hamiltonian constraint in terms of holonomies by means of Dirac observables. 

In particular, we are able to completely determine the quantum tomograms corresponding to both coherent states and the cat superposition of states considered as fiducial quantum states for the FRW Universe.   As we will see below, the obtained tomograms are not just consequent with Heisenberg's uncertainty relations but allow us to recover in a simple manner the expectation value of the volume operator as discussed within the quantum bounce scenario in the simplified representation of LQC.

The rest of the paper is organized as follows.  In Section~\ref{sec:DynLQC}, we briefly outline some relevant issues regarding the FRW model, in order to introduce some notation and to recall the corresponding eigenfunctions obtained within the LQC setup.   In Section~\ref{sec:Wigner}, we introduce the Weyl quantization map together with the Wigner quasi-probability function, putting special attention to the Wigner functions associated with a coherent Gaussian and with the so-called Schr\"odinger cat quantum 
states, respectively.  In section~\ref{sec:QT}, we analyze in detail the tomographic representation for the dynamics of the FRW model as viewed from the LQC perspective.  In particular, we discuss the obtained tomograms for the coherent and cat quantum states, respectively, and we calculate the expectation values for the volume operators in each case.  Finally, we include in Section~\ref{sec:conclu} some concluding remarks.

\section{Dynamics in Loop Quantum Cosmology}
\label{sec:DynLQC}
\subsection{Classical Dynamics}

In this section, we start by summarizing some key features related to the dynamics of the loop quantization of cosmological models (for more technical details, we refer the reader to \cite{Improved}, \cite{MLQC}, \cite{Robustness}, \cite{LQCSR}). For simplicity, we will focus our attention to study the quantization of the flat, isotropic and homogeneous Friedmann-Robertson-Walker (FRW) model in the presence of a massless scalar field. The metric in this model can be written as
\begin{equation}
ds^{2}=-N^{2}(x)dt^{2}+q_{ab}dx^{a}dx^{b},
\end{equation}
where $N(x)$ denotes the lapse function and $q_{ab}$ entails the spatial part of the metric expressed as
\begin{equation}
q_{ab}=\delta_{ij}\omega^{i}_{a}\omega^{j}_{b}=a^{2}(t){}^{o}q_{ab}=a^{2}(t)\delta_{ij}{}^{o}\omega^{i}_{a}{}^{o}\omega^{j}_{b},
\end{equation} 
where the function $a(t)$ stands for the scale factor, while $\{ {}^{o}\omega^{i}_{a}$, ${}^{o}e^{a}_{i}\}$ are a set of orthonormal cotriads and triads compatible with the fiducial flat metric ${}^{o}q_{ab}$, satisfying ${}^{o}\omega^{i}_{a}\,{}^{o}\e^{a}_{j}=\delta^{i}_{j}$.
Within this cosmological setup, the classical dynamics is defined by the Hamiltonian constraint
\begin{equation}\label{total}
H=H_{grav}+H_{\phi}\approx 0, 
\end{equation}
where the gravitational part for the flat FRW model reads (see, e.g. \cite{Improved}, \cite{MLQC})
\begin{equation}\label{Hg}
H_{grav}=-\gamma^{-2}\int_{\mathcal{V}}d^{3}x\;Ne^{-1}\varepsilon_{ijk}E^{ai}E^{bj}F^{k}_{ab}.
\end{equation}
In this expression, $\mathcal{V}$ describes an elementary cell embedded in a spacelike hypersurface with fiducial volume $V_{0}=\int_{\mathcal{V}}d^{3}x\sqrt{{}^{o}q}$.  Also, $\gamma$ denotes the Barbero-Immirzi parameter, and $\varepsilon_{ijk}$ represents the alternating tensor, while $E^{a}_{i}$ is a densitized vector field 
for which $e=\sqrt{|\det E|}$.  Finally, $F_{ab}^{k}$ stands for the curvature associated with the $SU(2)$ connection $A^{k}_{a}$.	

The description of the quantum theory within the LQC framework is established by writing the curvature $F^{k}_{ab}$, also called the field strength, in terms of holonomy variables around loops \cite{Improved}, just as it is the case for 
non-Abelian gauge theories,
\begin{equation}\label{curvature}
F_{ab}^{k}=-2\lim_{Ar\square\to 0}\tr\left( \frac{h_{\square_{ij}}^{(\mu)}-1}{\mu^{2}V_{0}^{2/3}}\right)\tau^{k} \, {}^{o}\omega^{i}_{a} \, {}^{o}\omega^{j}_{b},
\end{equation} 
where $h_{\square_{ij}}^{(\mu)}$ indicates the holonomy around the square loop $\square_{ij}$ spanned by a face of the elementary cell with area $Ar\square$ and 
whose individual sides have length $\mu V_{0}^{1/3}$ with respect to the fiducial metric ${}^{o}q_{ab}$.  Thus, we may decompose  $h_{\square_{ij}}^{(\mu)}$  as the product of holonomies all over the four edges of the square loop $\square_{ij}$, that is,
\begin{equation}
h_{\square_{ij}}^{(\mu)}=h^{(\mu)}_{i}h^{(\mu)}_{j}(h^{(\mu)}_{i})^{-1}(h^{(\mu)}_{j})^{-1} \,.
\end{equation}
Further, the holonomy  $h^{(\mu)}_{k}$ of the connection $A^{i}_{a}$ along the line segment in the direction ${}^{o}e^{a}_{k}$ with length $\mu V_{0}^{1/3}$ reads
\begin{equation}\label{holonomy}
h^{(\mu)}_{k}(c)=\cos(\mu c/2)I+2\sin(\mu c/2)\tau_{k},
\end{equation}
where $I$ denotes the unit $2\times 2$ matrix, and $\tau_{k}=-i\sigma_{k}/2$ stand for the basis elements of the $su(2)$ Lie algebra, written in term of the Pauli matrices $\sigma_{k}$. Considering that we are dealing with spatially flat, isotropic models, the gravitational phase space variables, namely the connection $A^{i}_{a}$ and the 
density-weighted triads $E^{a}_{i}$, may be explicitly determined as
\begin{equation}
A^{i}_{a}=c  \, {}^{o}\omega^{i}_{a}V_{0}^{-1/3}, \;\;\;\;\textup{and}\;\;\;\; E^{a}_{i}=p \, {}^{o}e^{a}_{i}\sqrt{{}^{o}q}V_{0}^{-2/3},
\end{equation}
where $c=\gamma\dot{a}V_{0}^{1/3}$ and $|p|=a^{2}V_{0}^{2/3}$, are the symmetry reduced phase space variables satisfying the Poisson algebra $\left\lbrace c,p \right\rbrace=8\pi G\gamma/3 $, which may be demonstrated to be invariant under the choice of any fiducial metric~\cite{Improved},~\cite{Robustness}.

In analogy to the full LQG scenario, and by considering the very useful identity 
\begin{equation}
e^{i}_{a}=\frac{1}{4\pi G\gamma}\left\lbrace A^{i}_{a},V\right\rbrace, 
\end{equation}
the term comprising the triads in equation (\ref{Hg}) can be expressed as
\begin{equation}\label{triads}
\varepsilon_{ijk}e^{-1}E^{ai}E^{bj}=\frac{\sgn(p)}{2\pi G\gamma\mu V_{0}^{1/3}}\sum_{k}{}^{o}\varepsilon^{abc}\,{}^{o}\omega^{k}_{c}\tr\left( h^{(\mu)}_{k}\left\lbrace (h^{(\mu)}_{k})^{-1},V\right\rbrace \tau_{i}\right) ,
\end{equation}
where $V=|p|^{3/2}=a^{3}V_{0}$ denotes the volume of the elementary cell $\mathcal{V}$, this formula holds for any choice of $\mu$ \cite{Thiemann}. By combining the expressions (\ref{curvature}) and (\ref{triads}), the gravitational part of the total Hamiltonian reads \cite{Improved},
\begin{equation}
H_{grav}=\lim_{\mu\to 0}H_{grav}^{(\mu)},
\end{equation} 
where we have defined
\begin{equation}
\mkern-50mu H^{(\mu)}_{grav}=-\frac{\sgn(p)}{2\pi G\gamma^{2}\mu^{3}}\sum_{ijk}N\varepsilon^{ijk}\tr\left( h^{(\mu)}_{i}h^{(\mu)}_{j}(h^{(\mu)}_{i})^{-1}(h^{(\mu)}_{j})^{-1}h^{(\mu)}_{k}\left\lbrace(h^{(\mu)}_{k})^{-1},V \right\rbrace \right). 
\end{equation}
In addition, the classical Hamiltonian corresponding to the massless scalar field can be written in terms of the gravitational conjugate variables $(c,p)$ as follows 
\begin{equation}
H_{\phi}=\frac{N}{2}p_{\phi}^{2}|p|^{-3/2},
\end{equation}
where $\phi$ and $p_{\phi}$ satisfy the Poisson commutation relation $\left\lbrace\phi,p_{\phi} \right\rbrace=1$. Then, following the general considerations stated in \cite{Perez}, the substitution of equation (\ref{holonomy}) into the expression that determines $H^{(\mu)}_{grav}$, allows us to obtain the total Hamiltonian constraint (\ref{total}) within the LQC framework as
\begin{equation}\label{HLQC}
H^{(\lambda)}=N\left(-\frac{3}{8\pi G\gamma^{2}}\frac{\sin^{2}(\lambda\beta)}{\lambda^{2}}v+\frac{p^{2}_{\phi}}{2v} \right)\approx 0, 
\end{equation}
where, following~\cite{Improved}, \cite{Robustness}, we have defined a new set of canonical variables
\begin{equation}
\beta:=\frac{c}{|p|^{1/2}}\,, \hspace{5ex} v:=|p|^{3/2}\,.
\end{equation}
In view of the quantum nature of geometry inherent to the loop quantization program, the parameter $\lambda:=\mu|p|^{1/2}=\mu aV_{0}^{1/3},$ corresponds to a physical length whose value is related to the minimum eigenvalue of the area operator $\Delta=(2\sqrt{3}\pi\gamma)l_{P}^{2}$ in LQG through the constraint $\lambda^{2}=\Delta$ and, in principle, it may be fixed by cosmological data \cite{Improved}, \cite{MLQC}. The canonical variables $(\beta,v)$ amount to terms  proportional to the Hubble parameter $\beta=\gamma\dot{a}/a$ and the scale factor $v^{1/3}=aV_{0}^{1/3}$, respectively.

In what follows, our strategy is to apply the Reduced Phase Space (RPS) approach of LQC (see, e.g., \cite{EnergyBB},  \cite{TurningBB}, \cite{Algebraic}, \cite{Length}, \cite{Bianchi}) to the Hamiltonian constraint depicted in (\ref{HLQC}). Contrary to the standard LQC scenario, where the Hamiltonian $H^{(\lambda)}$ (with the parameter $\lambda$ taken different from zero) is promoted to an operator acting on the Bohr representation for almost periodic functions, and then imposing the total Hamiltonian 
constraint (\ref{HLQC}) only at the quantum level \cite{LQCSR}, within the RPS approach the point of departure is to solve the modified Hamiltonian constraint in terms of holonomies at the classical level, and then, we may identify the physical phase space by means of Dirac's observables \cite{TBB2} which, subsequently, are employed to establish a quantum representation on an appropriate Hilbert space. Both methods provide similar results, in particular those related to the introduction of big bounce scenarios and to the spectrum properties of the volume operator.
Nevertheless, the RPS approach allows us to derive analytical expressions which will be useful later in order to obtain the corresponding quantum tomograms by means of the Radon integral transformation.

A function $\mathcal{O}$ is called a Dirac observable on the phase space spanned by the canonical variables $(\beta,v,\phi, p_{\phi})$ if 
\begin{equation}
\left\lbrace\mathcal{O},H^{(\lambda)} \right\rbrace\approx 0, 
\end{equation}
where the Poisson bracket is defined to be
\begin{equation}
\left\lbrace\cdot,\cdot \right\rbrace=4\pi G\gamma\left(\frac{\partial}{\partial\beta}\frac{\partial}{\partial v}-\frac{\partial}{\partial v}\frac{\partial}{\partial\beta} \right)+\left( \frac{\partial}{\partial\phi}\frac{\partial}{\partial p_{\phi}}-\frac{\partial}{\partial p_{\phi}}\frac{\partial}{\partial \phi}\right).   
\end{equation}
Following the construction stated in \cite{TurningBB}, one may realize that there are only two Dirac observables which parameterize the physical phase space
\begin{equation}
\mathcal{O}_{1}=p_{\phi}, \hspace{5ex} \mathcal{O}_{2}=\phi-\frac{\sgn(p_{\phi})}{3\kappa}\, \arcoth(\cos(\lambda\beta)).
\end{equation}
where $\kappa^{2}=4\pi G/3$. This implies that even though our kinematic phase space is four dimensional, the solution of the total Hamiltonian constraint 
(\ref{HLQC}) renders the physical phase space two dimensional.  Finally, if one introduce the variables
\begin{equation}\label{volume}
q:=\beta \,, \hspace{5ex} p:=\frac{1}{4\pi G\gamma}v \,, \hspace{5ex} t:=-\sgn(p_{\phi})\sqrt{3\pi}G\phi \,,
\end{equation}
it can be proved \cite{TurningBB} that the system evolves by means of the physical Hamiltonian
\begin{equation}\label{Ham}
H_{\lambda}=\frac{2}{\lambda\sqrt{G}}p\sin(\lambda q),
\end{equation}
where a particular gauge fixing for the lapse function $N$ has been selected in order to simplify the calculations, given that relative dynamics is gauge independent.

In this manner, and according to Dirac's prescription for constrained systems \cite{Dirac}, one can assert that the dynamical initial constraint has been solved. Within the Hamiltonian (\ref{Ham}), the variable $t$ corresponds to the intrinsic time for the relative dynamics in terms of the scalar field $\phi$, and the variable $p$ results to be proportional to the total volume of space provided that the topology is compact. Furthermore, $H_{\lambda}$ is positive definite if $q\in (0,\pi/\lambda)$ and $p>0$, enabling us to remove the appearance of possible dynamical instabilities.

\subsection{Quantum dynamics}

In order to perform the quantization of the system, let us introduce the Hilbert space $\mathcal{H}=L^{2}([0,\pi/\lambda],dq)$, in this manner, the position and momentum operator act on any $\varphi\in\mathcal{H}$ as usual (in natural units where $\hbar=1$)
\begin{equation}\label{posmom}
\hat{q}\varphi(q)=q\varphi(q), \;\;\;\; \hat{p}\varphi(q)=-i\frac{d}{dq}\varphi(q).
\end{equation}
Since the Hamiltonian $H_{\lambda}$ is positive definite if $q\in (0,\pi/\lambda)$ and $p>0$, the corresponding quantum Hamiltonian is given by the symmetric operator
\begin{equation}
\hat{H}_{\lambda}=\frac{2}{\lambda\sqrt{G}}\widehat{\sin^{1/2}(\lambda q)}\,\hat{p}\,\widehat{\sin^{1/2}(\lambda q)}
\end{equation}
which proves to be equivalent to the normal ordered expression \cite{QSBU}
\begin{equation}
\hat{H}_{\lambda}=\frac{1}{\lambda\sqrt{G}}\left( \hat{p}\,\widehat{\sin(\lambda q)}+\widehat{\sin(\lambda q)}\,\hat{p}\right). 
\end{equation}
By means of the Schr\"odinger representation of the position and momentum operators given in (\ref{posmom}), the resulting Hamiltonian operator reads
\begin{equation}\label{Hhat}
\hat{H}_{\lambda}\varphi=-\frac{i}{\lambda\sqrt{G}}\left( 2\sin(\lambda q)\frac{d}{dq}+\lambda\cos(\lambda q)\right)\varphi, 
\end{equation}
where $\varphi\in D(\hat{H}_{\lambda})\subset\mathcal{H}$, and $D(\hat{H}_{\lambda})$ denotes a densely defined domain of the operator $\hat{H}_{\lambda}$.
The equation associated with the Hamiltonian operator thus takes the explicit 
eigenvalue form
\begin{equation}
-\frac{i}{\lambda\sqrt{G}}\left( 2\sin(\lambda q)\frac{d\varphi}{dq}+\lambda\cos(\lambda q)\varphi\right)=E\varphi, 
\end{equation}
where $E\in\mathbb{R}$. The corresponding eigenfunctions are given by
\begin{equation}\label{eigenfunction}
\varphi_{E}=\frac{C}{\sqrt{\sin(\lambda q)}}\exp\left\lbrace \frac{i}{2}\sqrt{G}E\ln \left|\tan\left( \frac{\lambda q}{2}\right)\right|\right\rbrace, 
\end{equation}
where $C=(\lambda\sqrt{G}/4\pi)^{1/2}$ is a normalization constant such that the solutions satisfy the orthonormality condition
\begin{equation}
\braket{\varphi_{E},\varphi_{E'}}=\delta(E-E').
\end{equation}
Furthermore, according to the von Neumann's theorem \cite{Reed}, it is shown that the symmetric operator $\hat{H}_{\lambda}$, also defines a self-adjoint operator on the dense domain \cite{Evolution}, specified by 
\begin{equation}
D(\hat{H}_{\lambda})=\spann\left\lbrace \eta_{n}, n\in\mathbb{Z}\right\rbrace ,
\end{equation}
where 
\begin{equation}
\eta_{n}(q)=\int_{-\infty}^{\infty}dE\,f_{n}(E)\varphi_{E}(q), \;\;\;\textup{for}\;\;\;f_{n}(E)\in C^{\infty}_{0}(\mathbb{R}).
\end{equation}
This means, by means of the spectral theorem, that the spectrum of $\hat{H}_{\lambda}$ belongs to the real axis  (other properties related to this Hamiltonian operator, but in the context of the zeros of the Riemann zeta function, can be found in \cite{Riemann}).

In the next section, our purpose is to determine the Wigner function and the quantum tomogram associated with the FRW model within the LQC framework. Since the eigenfunctions (\ref{eigenfunction}) are confined to a finite interval, in order to obtain the Wigner function, the Hamiltonian given in (\ref{Ham}) should be complemented by introducing an infinite barrier potential. However, the implementation of this procedure is rather cumbersome as shown in \cite{Dias}, \cite{Prata}. To surpass such difficulty, let us consider the mapping $\mathcal{U}:L^2([0,\pi/\lambda],dq)\to L^{2}(\mathbb{R},dQ)$, defined by
\begin{equation}\label{U}
\mathcal{U}\varphi(q)=\sqrt{\frac{2}{\lambda\sqrt{G}}}\sqrt{\sin(\lambda q)}\varphi(q):=\psi(Q),
\end{equation}      
where $\varphi(q)\in L^2([0,\pi/\lambda],dq)=\mathcal{H} $ and $\psi(Q)\in L^{2}(\mathbb{R},dQ)=\mathcal{H}'$. By using this transform, the eigenfunctions $\varphi_{E}(q)$ are mapped into plane waves \cite{Robustness}, \cite{QSBU} as
\begin{equation}
\mathcal{U}:\varphi_{E}(q)\to \psi_{E}(Q)=\frac{1}{\sqrt{2\pi}}e^{iEQ}.
\end{equation}
This map also defines an isometry and hence, a unitary transform between the Hilbert spaces $\mathcal{H}$ and $\mathcal{H}'$, since
\begin{eqnarray}
\braket{\varphi_{1},\varphi_{2}}_{\mathcal{H}}&=&\int_{0}^{\pi/\lambda}dq\,\overline{\varphi_{1}(q)}\varphi_{2}(q), \nonumber \\
&=& \int_{-\infty}^{\infty}dQ\,\overline{\psi_{1}(Q)}\psi_{2}(Q), \nonumber\\
&=& \braket{\psi_{1},\psi_{2}}_{\mathcal{H}'}.
\end{eqnarray}
Finally, under this unitary transform, it can be proved that the Hamiltonian operator, $\hat{H}_{\lambda}\varphi(q)$, acting on $\varphi\in\mathcal{H}$, is mapped into the operator $\hat{P}\psi(Q)=-id\psi(Q)/dQ$, acting on $\psi(Q)\in\mathcal{H}'$, as
\begin{equation}
\hat{P}\psi(Q)=\mathcal{U}^{-1}\hat{P}\mathcal{U}\varphi(q)=\hat{H}_{\lambda}\varphi(q).
\end{equation}
Consequently, the time evolution given in the Hilbert space $\mathcal{H}$  
merely corresponds to a translation in the Hilbert space $\mathcal{H}'$, since
\begin{eqnarray}\label{Evolution}
\hat{U}\varphi(q):=e^{-i\hat{H}_{\lambda}t}\varphi(q)=e^{-i\hat{P}t}\psi(Q)=\psi(Q-t),
\end{eqnarray}
this means that the shape acquired by probability distributions are preserved in time, and the quantum dynamics results in shifting the initial state in the position variable $Q$, as discussed in detail in~\cite{QSBU}.

\section{Phase space Quantum Mechanics}
\label{sec:Wigner}
\subsection{Weyl quantization and Wigner function}

The general quantization procedure, know as Weyl quantization \cite{Weyl}, \cite{Moyal}, is based on the construction of a map between real-valued functions, defined on the classical phase space, and self-adjoint operators acting on a Hilbert space, such that, the algebraic properties of classical functions are preserved by the quantum operators. Let us consider $f\in \mathcal{S}'(\mathbb{R}^{2})$ an arbitrary function on the phase space $\mathbb{R}^{2}$, where $\mathcal{S}'(\mathbb{R}^{2})$ stands for the Schwartz space of tempered distributions, this means, the space defined by continuous linear functionals on the Schwartz space of rapidly decreasing smooth functions, $\mathcal{S}(\mathbb{R}^{2})$, \cite{Reed II}. We define the Weyl quantization map of $f$, as an operator acting on the Hilbert space $L^{2}(\mathbb{R},dQ)=\mathcal{H}'$, by
\begin{equation}\label{Weyl}
\mathcal{Q}(f)=\frac{1}{4\pi^{2}}\int_{\mathbb{R}^{2}}d\lambda d\xi \tilde{f}(\lambda,\xi)e^{i(\lambda \hat{Q}+\xi\hat{P})},
\end{equation}
where $\tilde{f}(\lambda,\xi)$ denotes the Fourier transform,
\begin{equation}\label{Fourier}
\tilde{f}(\lambda,\xi)=\int_{\mathbb{R}^{2}}dQ dP f(Q,P)e^{-i(\lambda Q+\xi P)},
\end{equation}
and the operators $\hat{Q}$, $\hat{P}$ satisfy the canonical commutator relations
\begin{eqnarray}
\left[ \hat{Q},\hat{P}\right] \psi(Q)&=& i\psi(Q) \,, \\
\left[ \hat{Q},\hat{Q}\right]\psi(Q) &=& 0=[\hat{P},\hat{P}]\psi(Q) \,,
\end{eqnarray}
for $\psi(Q)\in\mathcal{H}'$. By substituting the expression of the Fourier transform (\ref{Fourier}) into (\ref{Weyl}), the Weyl quantization map associated with the function $f$ reads
\begin{equation}
\hat{f}=\mathcal{Q}(f)=\frac{1}{2\pi}\int_{\mathbb{R}^{2}}dQ dP f(Q,P)\hat{\Delta}(Q,P),
\end{equation}
where the integral operator $\hat{\Delta}(Q,P)$ is given by
\begin{equation}
\hat{\Delta}(Q,P)=\frac{1}{2\pi}\int_{\mathbb{R}^{2}}d\lambda d\xi e^{-i\lambda(Q-\hat{Q})-i\xi (P-\hat{P})}. 
\end{equation}
The integral operator $\hat{\Delta}(Q,P)$ corresponds to the so called Weyl-Stratonovich quantizer \cite{Stratonovich}, and satisfies the following properties
\begin{eqnarray}
\hat{\Delta}^{\dagger}(Q,P)=\hat{\Delta}(Q,P), \\
\tr\left\lbrace \hat{\Delta}(Q,P)\right\rbrace =1, \\
\tr \left\lbrace \hat{\Delta}(Q,P)\hat{\Delta}(Q',P')\right\rbrace =2\pi\delta(Q-Q')\delta(P-P'), \label{consistency}
\end{eqnarray}  
Now, given an operator $\hat{f}$ acting on the Hilbert space $\mathcal{H}'$, and by making use of the relation (\ref{consistency}), also known as the compatibility condition, we can define its corresponding Weyl symbol \cite{Folland}, as  
\begin{equation}
W_{\hat{f}}(Q,P):=\tr\left\lbrace \hat{f}\hat{\Delta}(Q,P)\right\rbrace, 
\end{equation}
which can be understood as the inverse relation of the Weyl quantization map, since $\mathcal{Q}(W_{\hat{f}}(Q,P))=\hat{f}$.
For the specific case of the density operator $\hat{\rho}=\ket{\psi}\bra{\psi}$, associated with a quantum state $\psi\in\mathcal{H}'$, its Weyl symbol reads
\begin{eqnarray}\label{Wigner}
\rho(Q,P)&:=& W_{\hat{\rho}}(Q,P)=\tr\left\lbrace \hat{\rho}\hat{\Delta}(Q,P)\right\rbrace, \nonumber \\
&=&\frac{1}{2\pi}\int_{\mathbb{R}}dY\bar{\psi}\left(Q+\frac{Y}{2}\right)\psi\left(Q-\frac{Y}{2} \right)e^{iPY}.   
\end{eqnarray}
This real-valued function, defined on the phase space $\mathbb{R}^{2}$, is called the Wigner function associated with the Hilbert space $\mathcal{H}'$. As one may easily check \cite{Curtright}, it is normalized 
\begin{equation}
\int_{\mathbb{R}}dQ\,dP\,\rho(Q,P)=1,
\end{equation}
and the projections onto the positions and momentum variables leads to the marginal probability densities
\begin{eqnarray}
\int_{\mathbb{R}}dQ\,\rho(Q,P)=|\psi(P)|^{2}, \\
\int_{\mathbb{R}}dP\,\rho(Q,P)=|\psi(Q)|^{2}.
\end{eqnarray}
In addition, by employing the Cauchy-Schwartz inequality, $\rho(Q,P)$ results to be bounded
\begin{equation}
-\frac{1}{\pi}\leq \rho(Q,P)\leq \frac{1}{\pi},
\end{equation} 
which means that the Wigner function for a quantum state is allowed to acquire negative values on certain regions of the phase space. Therefore, it can not be merely interpreted as a probability distribution in the sense of standard statistical mechanics and, as a consequence, it is usually referred to as a quasi-probability distribution. However, this apparently odd feature of the Wigner function has
been proved to be a useful tool as an indicator of quantumness by measuring correlations and coherence effects between quantum states, such as squeezing and superposition, precisely by means of its negative values \cite{Negativity}.
Moreover, the Wigner function can be used to calculate the expectation value of any operator by integrating its corresponding Weyl symbol over the phase space, 
\begin{equation}
\braket{\hat{f}}:=\braket{\psi,\hat{f}\psi}=\int_{\mathbb{R}^{2}}dQ\,dP\,\rho(Q,P)W_{\hat{f}}(Q,P).
\end{equation}
All these properties suggest, that the Wigner function represents in phase space the closest object to define a probability distribution for a quantum system and,
similarly, the Weyl map corresponds to the quantum analogue of the characteristic generating functions in classical probability theory.

\subsection{Wigner functions for the LQC model} 

In Section 2.2 above we addressed the analysis of the quantum dynamics associated with the Friedmann-Robertson-Walker model in the presence of a massless scalar field within the Reduce Phase Space approach of LQC. As mentioned before, by means of the unitary transformation $\mathcal{U}$ stated in (\ref{U}), we were allowed to map the quantum physical Hamiltonian $\hat{H}_{\lambda}$, defined on the Hilbert space $\mathcal{H}=L^{2}([0,\pi/\lambda],dq)$, to the standard momentum operator $\hat{P}$, acting on the Hilbert space $\mathcal{H'}=L^{2}(\mathbb{R},dQ)$. This means, that the eigenvalue problem $\hat{P}\Psi_{P}=P\Psi_{P}$, has as solution, 
\begin{equation}\label{Psi}
\Psi(Q)=\int_{\mathbb{R}}dP\,c(P)\psi_{P}(Q),
\end{equation}   
where $\psi_{P}(Q)$ is an eigenstate of the $\hat{P}$ operator, and the wave packet profile $c(P)$ is represented as follows
\begin{equation}
c(P)=e^{-\alpha(P-P_{0})^{2}}.
\end{equation} 
The reason to focus on Gaussian states, centered in $P_{0}$ with dispersion $\alpha$, lies on the possibility to describe the semiclassical behavior of minisuperspace models of the Universe by means of decoherence between relevant degrees of freedom, such as the scale factor, and irrelevant degrees of freedom, such as those associated with perturbations \cite{Gaussian}, \cite{Numerical}, \cite{Decoherence}. Furthermore, even if the initial state is not Gaussian, the corresponding Wigner function can be approximated with a Gaussian distribution  by taking the quantum moments that describe the quantum system, up to quadratic order. 

By using expression (\ref{Wigner}) and the states with Gaussian profile (\ref{Psi}), the Wigner function reads
\begin{equation}\label{Wgaussian}
\rho(Q,P)=\frac{1}{\pi}e^{-\frac{Q^{2}}{2\alpha}-2\alpha(P-P_{0})^{2}},
\end{equation}
which, as we can observe, results to be a positive function, exhibiting the semiclassical behavior of the state. Moreover, according to Hudson's theorem \cite{Hudson}, this characteristic proves to be equivalent to the definition of a coherent state.
 
In addition, let us consider now a superposition of two Gaussian packets centered at $\pm P_{0}$. This configuration, known as a Schr\"odinger cat state, allows us to analyze decoherence in non-semiclassical situations which, within the LQC scenario
of our interest, may be related to a Universe in a superposition of two different orientation of the triads \cite{QSBU}, \cite{Kiefer}. The wave function corresponding to a cat state is given by
\begin{equation}\label{cats}
\Psi(Q)=\frac{B_0}{\sqrt{2}(2\pi\alpha)^{1/4}}\left( e^{-\frac{Q^{2}}{4\alpha}+iP_{0}Q}+e^{-\frac{Q^{2}}{4\alpha}- iP_{0}Q}\right), 
\end{equation}
where the normalization factor $B_0$ occurs to be
\begin{equation}\label{nfactor}
B_0=\frac{1}{\sqrt{1-e^{-2P^{2}_{0}\alpha}}},
\end{equation}
which means that the two Gaussian wave functions are not orthogonal. By substituting (\ref{cats}) into the definition of the Wigner function (\ref{Wigner}), the result is
\begin{equation}\label{Wcat}
\rho(Q,P)=\frac{B_0^{2}}{\pi}e^{-\frac{Q^{2}}{2\alpha}-2\alpha P^{2}}\left[\cos(2P_{0}Q)+\cosh(4\alpha P_{0}P)e^{-2\alpha P^{2}_{0}} \right]. 
\end{equation}
The appearance of the oscillatory cosine term represents the interference between the two Gaussian wave packets. Evidently, this Wigner function takes both positive and negative values, and then, it cannot be interpreted as a probability distribution in the sense of classical statistical mechanics, as mentioned above. Furthermore, as $P_{0}\to\infty$ the amplitude of the interference pattern does not diminish, implying that the Wigner function associated with a cat state is not reduced to a classical probability distribution in the macroscopic limit. Finally, since the evolution operator corresponds to translations as stated in (\ref{Evolution}), the linear properties of the Wigner function in phase space \cite{Curtright}, allow us to determine straightforwardly the time dependent Wigner functions associated with the Gaussian state and the cat state, by substituting the position variable $Q$ with $Q-t$, in the expressions (\ref{Wgaussian}) and (\ref{Wcat}), respectively.

\section{Quantum Tomography}
\label{sec:QT}

In this section, we will consider the quantum dynamics corresponding to the FRW model in the LQG scheme, within the tomographic representation of quantum mechanics.

\subsection{The Tomographic representation}

As a first step, let us consider the quadrature associated with the quantum observable $\hat{X}$, which corresponds to a generic linear combination of the position and momentum operators $\hat{Q}$ and $\hat{P}$ of the form  
\begin{equation}
\hat{X}=\chi \hat{Q}+\nu\hat{P},
\end{equation}
where $\chi, \nu\in\mathbb{R}$, denote real numbers labeling an ensemble of rotated and scaled reference frames in the phase space at which the observable $\hat{X}$ is measured \cite{Ariano}.  Given $\hat{\rho}$, the density operator associated with a quantum system,  we define the characteristic function $G(\zeta)$, as the mean value of the exponential of the operator $\hat{X}$
\begin{equation}\label{mean}
G(\zeta):=\braket{e^{i\zeta\hat{X}}}=\tr\left\lbrace \hat{\rho}e^{i\zeta\hat{X}}\right\rbrace. 
\end{equation}
Then, by applying the Fourier transform to the characteristic function $G(\zeta)$, we obtain
\begin{equation}\label{TFourier}
w(X,\chi,\nu):=\frac{1}{2\pi}\int_{\mathbb{R}}d\zeta\,G(\zeta)e^{-i\zeta X}.
\end{equation}
The function $w(X,\chi,\nu)$ constitutes a marginal distribution as established in \cite{Manko}, \cite{Tombesi} and, as we shall observe, it completely characterizes the state of a quantum system by means of genuine probability distributions. By writing the trace of the density operator $\hat{\rho}$ in terms of the Wigner function, as defined in (\ref{Wigner}), the function $w(X,\chi,\nu)$ reads
\begin{equation}\label{Radon}
w(X,\chi,\nu)=\frac{1}{2\pi}\int_{\mathbb{R}^{3}}d\zeta\,dQ\,dP\,\rho(Q,P)e^{-i\zeta(X-\chi Q-\nu P)}. 
\end{equation}
In a general sense, the integral transform of the Wigner function stated in (\ref{Radon}), corresponds to a Radon transform known as the symplectic tomogram. The reason of this terminology follows from the fact that the operator $\hat{X}$ together with its conjugate $\hat{Y}=-s^{2}\nu Q+s^{-2}\chi P$, satisfies the canonical commutation relations $[X,Y]=[Q,P]=i$, where $s$ is a scaling parameter, such that $\chi=s\cos\theta$ and $\nu=s\sin\theta$, with an orientation labeled by $\theta$. This implies that the observable $\hat{X}$ yields a new position operator after a quantum canonical transformation. By evaluating the expectation value (\ref{mean}) through eigenstates of the operator $\hat{X}$, that is, $\ket{X}$ such that $\hat{X}\ket{X}=X\ket{X}$, it can be verified that the symplectic tomogram is proportional to $\braket{X,\hat{\rho}X}$, which means that  $w(X,\chi,\nu)$ results in a positive and normalized function, i.e., a probability distribution, since
\begin{equation}\label{normalization}
\int_{\mathbb{R}}dX\,w(X,\chi,\nu)=1.
\end{equation}
In case that $\chi=\cos\theta$ and $\nu=\sin\theta$, where the rotation angle $\theta$ labels the reference frame in classical phase space, the marginal distribution defined by the symplectic tomogram (\ref{Radon}) constitutes the homodyne distribution used in optical tomography \cite{Homodyne}. 
Formula (\ref{Radon}) can be inverted, allowing us to express the Wigner function in terms of the tomogram  as
\begin{equation}\label{WignerT}
\rho(Q,P)=\frac{1}{2\pi}\int_{\mathbb{R}^{3}}dX\,d\chi\,d\nu\,w(X,\chi,\nu)e^{-i(\chi Q+\nu P-X)}.
\end{equation}  
Considering that the Wigner function $\rho(Q,P)$ suffices to characterize the quantum state of a system and, on the other hand, the Wigner function is formulated in terms of the quantum tomogram, this means that the information associated with a quantum state is contained in $w(X,\chi,\nu)$, which consists in a probability distribution formed by an ensemble of rotated and squeezed quadratures.   

Within the symplectic tomographic scheme, and by using the approach developed in \cite{Manko2}, \cite{TQFT2}, the tomographic symbol associated with an operator $\hat{f}\in\mathcal{L}(\mathcal{H'})$, denoted by $\mathcal{T}_{\hat{f}}(Q,P)$, is obtained by means of the operator
\begin{equation}
\hat{\Delta}^{T}(X,\chi,\nu)=\delta(X\hat{1}-\chi\hat{Q}-\nu\hat{P}),
\end{equation}  
where $X,\chi,\nu\in\mathbb{R}$ and $\hat{1}$ denotes the identity operator. This means that 
\begin{equation}\label{Tsymbol}
\mathcal{T}_{\hat{f}}(Q,P)=\tr\left\lbrace\hat{f}\hat{\Delta}^{T}(X,\chi,\nu)\right\rbrace. 
\end{equation} 
In analogy to the Weyl quantization, the compatibility condition (\ref{consistency}) within the tomographic representation reads
\begin{equation}\label{Tcompatibility}
\tr\left\lbrace \hat{\Delta}^{T}(X,\chi,\nu)\mathcal{C}(X',\chi',\nu')\right\rbrace=\delta(X-X')\delta(\chi-\chi')\delta(\nu-\nu'), 
\end{equation}
where the operator $\hat{\mathcal{C}}(X,\chi,\nu)$ is given by
\begin{equation}
\hat{\mathcal{C}}(X,\chi,\nu)=\frac{1}{2\pi}e^{i(X\hat{1}-\nu\hat{P}-\chi\hat{Q})}.
\end{equation}
One can invert the relation (\ref{Tsymbol}) by using the compatibility condition (\ref{Tcompatibility}), so that
\begin{equation}\label{inverse}
\hat{f}=\int_{\mathbb{R}^{3}}dX\,d\chi\,d\nu\,f(Q,P)\hat{\mathcal{C}}(X,\chi,\nu).
\end{equation}
To finish this section, it is worthwhile to mention that the symplectic tomogram can be applied to calculate the expectation value of any quantum observable $\hat{f}\in\mathcal{L}(\mathcal{H}')$. By making use of the Wigner function $\rho(Q,P)$ in terms of the tomographic representation (\ref{WignerT}), and the relation (\ref{inverse}) between operators and its tomographic symbols, we obtain
\begin{equation}
\braket{\hat{f}}=\tr\left\lbrace\hat{\rho}\hat{f} \right\rbrace=\tr\left\lbrace \int_{\mathbb{R}^{3}}dX\,d\chi\,d\nu\,w(X,\chi,\nu)\hat{\mathcal{C}}(X,\chi,\nu)\hat{f}\right\rbrace , 
\end{equation}
which can be written as
\begin{equation}\label{Tmean}
\braket{\hat{f}}=\int_{\mathbb{R}^{3}}dX\,d\chi\,d\nu\,w(X,\chi,\nu)\mathcal{T}^{d}_{\hat{f}}(X,\chi,\nu), 
\end{equation}
where we have defined $\mathcal{T}^{d}_{\hat{f}}(X,\chi,\nu)$, as the dual tomographic symbol to the operator $\hat{f}$, given by
\begin{equation}\label{duals}
\mathcal{T}^{d}_{\hat{f}}(X,\chi,\nu)=\frac{1}{2\pi}\tr\left\lbrace \hat{f}e^{i(X\hat{1}-\nu\hat{P}-\chi\hat{Q})} \right\rbrace. 
\end{equation}
Therefore, and in close analogy with the Wigner approach, the expectation values of quantum observables are calculated by integrating the product of the quantum tomogram with the dual tomographic symbol associated with the observable. Nevertheless, contrary to the Wigner function, the tomographic scheme provides a genuine positive measurable probability distribution which, consistently, may be related with the classical counterpart of the system \cite{Manko5}.

\subsection{Tomography in LQC}

With the preceding calculations, we now determine the quantum tomograms corresponding to the coherent state and the cat superposition, as fiducial quantum states  of the FRW universe within the LQC framework. First, let us compute the Radon transform (\ref{Radon}) of the time dependent Wigner function associated with the Gaussian profile (\ref{Wgaussian}). We have, then
\begin{eqnarray}\label{TGaussian}
w(X,\chi,\nu,t)&=&\frac{1}{2\pi^{2}}\int_{\mathbb{R}^{3}}d\zeta\,dQ\,dP\,e^{-\frac{(Q-t)^{2}}{2\alpha}-2\alpha(P-P_{0})^{2}}e^{-i\zeta(X-\chi Q-\nu P)}, \nonumber \\
&=&\frac{1}{\sqrt{2\pi}\sigma_{X}}e^{-\frac{(X-\braket{\hat{X}})^{2}}{2\sigma^{2}_{X}}},
\end{eqnarray}
where the expectation value $\braket{\hat{X}}$ is given by $\braket{\hat{X}}=\chi\braket{\hat{Q}}+\nu\braket{\hat{P}}=\chi t+\nu P_{0}$, and $\sigma_{X}$ stands for the quantum dispersion of the variable $X$ in the coherent state, $\sigma_{X}=\sqrt{\braket{\hat{X}^{2}}-\braket{\hat{X}}^{2}}$, which results in
\begin{equation}
\sigma^{2}_{X}=\chi^{2}\sigma_{Q}+\nu^{2}\sigma_{P}+2\chi\nu\sigma_{QP}=\chi^{2}\alpha+\frac{\nu^{2}}{4\alpha}.
\end{equation}
Analogously, the quantum tomogram for the cat state (\ref{cats}) takes the form
\begin{equation}\label{Tcats}
\hspace{-1em}w(X,\chi,\nu,t)=\frac{B_0^{2}}{\sqrt{2\pi}\beta}e^{-\frac{X^{2}+\braket{\hat{X}}^{2}}{2\beta^{2}}}\left[\cosh\left(\frac{X\braket{\hat{X}}}{\beta^{2}} \right)+\cos\left(\frac{2\alpha\chi X\braket{\hat{X}}}{\nu\beta^{2}} \right)\right],
\end{equation}
where $\beta^{2}=\chi^{2}\alpha+\frac{\nu^{2}}{4\alpha}$, and $B_0$ is the normalization factor depicted in (\ref{nfactor}). We can observe that, contrary to the case of the Wigner function, these quantum tomograms define positive measurable probability distributions, as both satisfy the normalization condition (\ref{normalization}). This implies that the expressions obtained in (\ref{TGaussian}) and (\ref{Tcats}) completely characterize the quantum system within the symplectic tomographic scheme.

In order to further analyze the properties of the introduced states, let us determine the quantum moments, and the quantum dispersions of the position and momentum operators by means of the quantum tomograms. Since the method requires the dual tomographic symbol to evaluate the expectation value of an observable (\ref{duals}), we have that
\begin{eqnarray}
\mathcal{T}_{\hat{Q}}^{d}(X,\chi,\nu)&=&\frac{1}{2\pi}\tr\left\lbrace Q e^{i(X\hat{1}-\nu\hat{P}-\chi\hat{Q})}\right\rbrace  \nonumber \\
&=&\frac{1}{(2\pi)^{2}}\int_{\mathbb{R}^{2}}dQ\,dP\,Q e^{i(X-\chi Q-\nu P)} \nonumber \\
&=&ie^{iX}\delta^{\, \prime}(\chi)\delta(\nu) 
\end{eqnarray}
and, analogously, we obtain
\beq
\hspace{7.7ex}    \mathcal{T}_{\hat{P}}^{d}(X,\chi,\nu) & = & ie^{iX}\delta(\chi)\delta^{\, \prime}(\nu)
\,,
\\
\hspace{7ex}    \mathcal{T}_{\hat{Q}^{2}}^{d}(X,\chi,\nu) & = & -e^{iX}\delta^{\, \prime\prime}(\chi)\delta(\nu)
\,,
\\
\hspace{7ex}      \mathcal{T}_{\hat{P}^{2}}^{d}(X,\chi,\nu) & = & -e^{iX}\delta(\chi)\delta^{\, \prime\prime}(\nu) 
\,,
\\
\mathcal{T}_{\frac{1}{2}(\hat{Q}\hat{P}+\hat{P}\hat{Q})}^{d}(X,\chi,\nu) & = & -e^{iX}\delta^{\, \prime}(\chi)\delta^{\, \prime}(\nu).
\eeq
Thus, by using formula (\ref{Tmean}), the mean value of the position operator for the Gaussian tomogram (\ref{TGaussian}) associated with a coherent state reads,
\begin{eqnarray}
\braket{\hat{Q}}&=&\int_{\mathbb{R}^{3}}dX\,d\chi\,d\nu\,w(X,\chi,\nu)\mathcal{T}^{d}_{\hat{Q}}(X,\chi,\nu) \nonumber \\
&=&t.
\end{eqnarray}
In the same way, we determine that $\braket{\hat{P}}=P_{0}$, $\braket{\hat{Q}^{2}}=\alpha+t^{2}$, $\braket{\hat{P}^{2}}=\frac{1}{4\alpha}+P_{0}^{2}$ and $\braket{\frac{1}{2}(\hat{Q}\hat{P}+\hat{P}\hat{Q})}=tP_{0}$. This means, that the quantum dispersions take the values
\begin{equation}
\sigma^{2}_{Q}=\braket{\hat{Q}^{2}}-\braket{\hat{Q}}^{2}=\alpha \,, \hspace{5ex} 
\sigma^{2}_{P}=\braket{\hat{P}^{2}}-\braket{\hat{P}}^{2}=\frac{1}{4\alpha} 
\end{equation}
and, in addition, the covariance satisfies
\begin{equation}
C_{QP}=\frac{1}{2}\braket{\hat{Q}\hat{P}+\hat{P}\hat{Q}}-\braket{\hat{Q}}\braket{\hat{P}}=0.
\end{equation}
Using the above relations, it can be verified that the Heisenberg uncertainty relation is satisfied, since
\begin{equation}
\sigma^{2}_{Q}\sigma^{2}_{P}-C_{QP}^{2}\geq\frac{1}{4}.
\end{equation}

Now, let us consider the case in which the quantum tomogram $w(X,\chi,\nu,t)$, corresponds to a Schr\"odinger cat superposition of coherent states, as given in (\ref{Tcats}). In this circumstance, by employing equation (\ref{Tmean}), one can verify that the mean values, the quantum dispersions and the covariance are given by, $\braket{\hat{Q}}=t$, $\braket{\hat{P}}=0$, $\sigma^{2}_{Q}=B_0^{2}\alpha\left[ 1+e^{-2\alpha P_{0}^{2}}(1-4\alpha P_{0}^{2})\right]$, $\sigma_{P}^{2}=\frac{1}{4\alpha}\left[ e^{-2\alpha P_{0}^{2}}+(1+4\alpha P_{0}^{2})\right] $, and $C_{QP}=0$ respectively. From these expressions, the Heisenberg uncertainty relation follows
\begin{equation}
\sigma_{Q}^{2}\sigma_{P}^{2}-C_{QP}^{2}=\frac{1}{4}(1+\xi(\alpha P_{0}^{2}))\geq \frac{1}{4},
\end{equation}
since $\xi(\alpha P_{0}^{2})$ corresponds to a positive function as discussed in \cite{QSBU}. The obtained relations constitute an important characteristic of the quantum states within the tomographic description as the uncertainty relations may provide constraints on admissible tomographic probabilities, which in turn should be validated by experimental measurements as done within the quantum optics context in~\cite{Exp1} and~\cite{Exp2}.

Further, based on the properties of the tomographic distribution, one can calculate the expectation value of the Dirac observable associated with the volume operator, which due to (\ref{volume}), reads 
\begin{equation}
\hat{v}=4\pi\sqrt{G}\gamma\hat{p}. 
\end{equation}
By using the isometry mapping $\mathcal{U}$ defined in (\ref{U}), the volume operator can be written in terms of the position $\hat{Q}$ and momentum $\hat{P}$ operators \cite{TBB2}, as
\begin{equation}
\hat{v}=2\pi l^{3}_{P}\gamma\lambda\hat{P}\cosh\left( \frac{2\hat{Q}}{\sqrt{G}}\right). 
\end{equation}
The dual tomographic symbol corresponding to $\hat{v}$ is thus given by
\begin{eqnarray}
\mathcal{T}^{d}_{\hat{v}}(X,\chi,\nu)&=&\frac{1}{2\pi}\tr\left\lbrace \hat{v}e^{i(X\hat{1}-\nu\hat{P}-\chi\hat{Q})} \right\rbrace  \nonumber \\
&=&\pi l^{3}_{P}\gamma\lambda i e^{iX}\delta^{\, \prime}(\nu)\left[\delta\left(\chi-\frac{2i}{\sqrt{G}} \right)+\delta\left(\chi+\frac{2i}{\sqrt{G}} \right)   \right].
\end{eqnarray}
Then, by means of the formula (\ref{Tmean}) and the Gaussian tomogram (\ref{TGaussian}), we find the expectation value of the volume operator 
\begin{eqnarray}
\braket{\hat{v}}&=&\int_{\mathbb{R}^{3}}dX\,d\chi\,d\nu\,w(X,\chi,\nu)\mathcal{T}^{d}_{\hat{v}}(X,\chi,\nu) \nonumber \\
&=&2\pi l^{3}_{P}\gamma\lambda P_{0}e^{2\alpha/G}\cosh\left(\frac{2t}{\sqrt{G}}\right).
\end{eqnarray}
By substituting the time variable $t$ in terms of the scalar field $\phi$ as depicted in (\ref{volume}), our last expression reads
\begin{equation}
\label{eq:vol}
\braket{\hat{v}}=2\pi l^{3}_{P}\gamma\lambda P_{0}e^{2\alpha/G}\cosh\left(\sqrt{12\pi G}\phi\right),
\end{equation}
which coincides with the results presented in \cite{Robustness}, \cite{Evolution}, where the quantum bounce scenario is analyzed within the simplified representation of LQC. From~(\ref{eq:vol}) we realize that the minimum value admitted by the coherent Gaussian state reads
\begin{equation}
\braket{\hat{v}}_{min}=2\pi l^{3}_{P}\gamma\lambda P_{0}e^{2\alpha/G}.
\end{equation}
Besides, in the case of the Schr\"odinger cat state (\ref{Tcats}), the expectation value of the volume yields $\braket{\hat{v}}=0$, suggesting that the superposition of states describing different orientations of the triad cancel each other. However, as studied in \cite{Kiefer}, one may introduce decoherence by considering interaction with fermionic matter, which may be viewed as a way to single out a preferred orientation during the cosmic evolution.

\section{Conclusions}
\label{sec:conclu}

In this paper, we have studied the tomographic representation associated with the Wigner quasi-probability function for the FRW model as prescribed by the LQC scenario.   As mentioned above, the tomographic representation allows us to define a positive probability distribution for an arbitrary quantum state, which perfectly suits to analyze quantumness, entanglement and semiclassical conditions as discussed in the literature.   In our case of interest, we thus focused on Gaussian and Schr\"odinger cat states constructed from the eigenvalues of the quantum Hamiltonian operator for the FRW model.  Such states are labeled by a parameter that corresponds to a physical length which, in the loop quantization program, is related to the minimum eigenvalue of the area operator.   By considering the Phase Space quantum mechanics program, we are able to obtain their Wigner functions and, by implementing a Radon integral transform, we adequately establish the tomograms corresponding to either the coherent Gaussian or a superposition of quantum states.  These tomograms completely characterize the quantum FRW model within the symplectic tomographic representation and allow us to realize in a simpler manner certain attributes of the model of our interest.  
Particularly, we have demonstrated that Heisenberg's uncertainty relation straightforwardly ensue from the quantum dispersion of the rotated and squeezed quadrature operator $\hat{X}$.  Further, the properties of the tomographic distributions allowed us to recognize in a simple manner the expectation values of the Dirac observable held by the volume operator.  Indeed, for the coherent Gaussian state, we recover the expectation value for the volume operator as reported for the quantum bounce scenario in the simplified representation of LQC, while for the Schr\"odinger cat state the vanishing of the expectation value for the same operator portrayed the cancellation of superposed states with different orientations for the involved triads.  

Accordingly, we expect that the introduced tomographic representation for the FRW model in the LQC context, results relevant in order to discuss some other important issues, such as entanglement, tomographic entropy, semiclassical conditions and relative quantum fluctuations of observables, which may exhibit some imprints on the cosmic microwave background. Furthermore, the characterization of quantum states by means of the Radon transformations of the Wigner functions, may possibly lead to methods of direct experimental authentication for tomograms, thus taking advantage of the well-developed techniques in the areas of Quantum optics and  Quantum information theory. This will be done elsewhere.

\section*{Acknowledgments}
The authors would like to acknowledge financial support from CONACYT-Mexico
under the project CB-2017-283838.

\section*{References}

\bibliographystyle{unsrt}

\begin{thebibliography}{l}

\bibitem{Manko}Mancini S., Man'ko V.~I. and Tombesi P., \emph{Wigner function and probability distribution for shifted and squeezed quadratures}, Quantum Semiclass.~Opt.~{\bf 7} 615 (1995).

\bibitem{Tombesi}Mancini S., Man'ko V.~I. and Tombesi P., \emph{Symplectic tomography as classical approach to quantum systems}, Phys.~Lett.~A~{\bf 213} 1 (1996), \texttt{arXiv:quant-ph/9603002}.

\bibitem{Simoni}
Man'ko V.~I., Marmo G., Simoni A., Sudarshan E.~C.~G. and Ventriglia F., \emph{A tomographic setting for Quasi-distribution function}, Rep.~Math.~Phys. {\bf 61}
337 (2008), \texttt{arXiv:quant-ph/0604148v2}. 

\bibitem{Manko4}Ibort A., Man'ko V.~I., Marmo G., Simoni A. and Ventriglia F., \emph{An introduction to
the tomographic picture of quantum mechanics}, Phys.~Scr.~{\bf 79} 065013 (2009), \texttt{arXiv:0904.4439 [quant-ph]}.

\bibitem{Asorey}
Asorey M., Ibort A., Marmo G. and Ventriglia F., \emph{Quantum tomography twenty years later}, Phys.~Scr.~{\bf 90} 074031 (2015), \texttt{arXiv:1510.08140 [math-ph]}.


\bibitem{Phasespace}
Zachos C.~K., Fairlie D.~B. and Curtright T.~L., Quantum Mechanics ins Phase Space: An Overview with Selected Papers (Singapure, World-Scientific, 2005).

\bibitem{Bayen}
Bayen F., Flato M., Fronsdal C., Lichnerowicz A. and Sternheimer D., \emph{Deformation theory and quantization I \& II} Ann.~Phys. {\bf 111} 61 (1978); Ann.~Phys. {\bf 111} 111 (1978).

\bibitem{Bordemann}
Bordemann M., \emph{Deformation quantization: a survey}, J.~Phys.: Conf.~Ser. {\bf 103} 012002 (2008).

\bibitem{Manko3}Mancini S., Man'ko V.~I. and Tombesi P., \emph{Classical-like description of quantum
dynamics by means of symplectic tomography}, Found.~Phys.~{\bf 27} 801 (1997), \texttt{arXiv:quant-ph/9609026}.

\bibitem{Manko2}Man'ko M.~A., Man'ko V.~I. and Vilela-Mendes R., \emph{Tomograms and other transforms:
A unified view}, J.~Phys.~A:~Math.~Gen.~{\bf 34} 8321 (2001), \texttt{arXiv:math-ph/0101025}.

\bibitem{DAriano}
D'Ariano G.~M., Paris M.~G.~A. and Sacchi M.~F., \emph{Quantum tomography},
Adv.~Imaging Electron Phys. {\bf 128}  205--308 (2003),
\texttt{arXiv:quant-ph/0302028}.

\bibitem{Helsen}
Helsen J., Battistel J. and Terhal B.~M., \emph{Spectral quantum tomography}, npj Quantum Inf. {\bf 5} 74 (2019), \texttt{arXiv:1904.00177 [quant-ph]}.

\bibitem{TQFT}Man'ko M.~A., Man'ko V.~I. and Thanh N.~C., \emph{Tomographic-probability representation of the quantum scalar field}, J.~Russ.~Laser~Res.~{\bf 30} 1 (2009).

\bibitem{TQFT2}Berra-Montiel J. and Cartas-Fuentevilla R., \emph{Deformation quantization and the tomographic representation of quantum fields}, IJGMMP~{\bf 14} 2050207 (2020), \texttt{arXiv:2006.07688 [hep-th]}. 

\bibitem{Mendes} 
Man'ko V.~I. and Mendes R.~V., \emph{Lyapunov Exponent In Quantum Mechanics. A Phase-space Approach}, Physica {\bf D145} 330--348 (2000), \texttt{arXiv:quant-ph/0002049}.

\bibitem{Capozzielo}Capozziello S., Man'ko V.~I., Marmo G. and Stornaiolo C., \emph{A tomographic description for classical and quantum cosmological perturbations}, Phys.~Scr.~{\bf 80} 045901 (2009), \texttt{arXiv:0905.1244 [gr-qc]}.

\bibitem{Capozzielo2}Capozziello S., Man'ko V.~I., Marmo G. and Stornaiolo C., \emph{Tomographic representation of minisuperspace quantum cosmology and Noether symmetries}. Gen.~Relativ.~Gravit.~{\bf 40} 2627 (2008), \texttt{arXiv:0706.3018 [gr-qc]}.

\bibitem{Stornaiolo}Man'ko V.~I., Marmo G. and Stornaiolo, C., \emph{Radon transform of the Wheeler-De Witt equation and tomography of quantum states of the universe}, Gen.~Relativ.~Gravit.~{\bf 37} 99 (2005), \texttt{arXiv:gr-qc/0307084}.

\bibitem{Improved}Ashtekar A., Pawlowski T. and Singh P., \emph{Quantum Nature of the Big Bang: Improved dynamics}, Phys.~Rev.~{\bf D74} 084003 (2006), \texttt{arXiv:gr-qc/0607039}.

\bibitem{MLQC}Ashtekar A., Bojowald M. and Lewandowski J., \emph{Mathematical structure of loop quantum cosmology}, Adv.~Theor.~Math.~Phys.~{\bf 7} 233 (2003), \texttt{arXiv:gr-qc/0304074}.

\bibitem{Robustness}Ashtekar A., Corichi A. and Singh P., \emph{Robustness of key features of loop quantum cosmology}, Phys.~Rev.~{\bf D77} 024046 (2008), \texttt{arXiv:0710.3565 [gr-qc]}.

\bibitem{LQCSR}Ashtekar A. and Singh P., \emph{Loop Quantum Cosmology: A Status Report}, Class.~Quantum~Grav.~{\bf 28} 213001 (2011), \texttt{arXiv:1108.0893 [gr-qc]}.

\bibitem{Brief}Ashtekar A. and Bianchi E., \emph{A Short Review of Loop Quantum Gravity}, Rep.~Prog.~Phys. {\bf 84} 042001 (2021), \texttt{arXiv:2104.04394 [gr-qc]}.

\bibitem{Fewster} 
Fewster C.~J.  and Sahlmann H., \emph{Phase space quantization and Loop Quantum Cosmology: A
Wigner function for the Bohr-compactified real line}, Class.~Quantum Grav. {\bf 25} 225015 (2008), \texttt{arXiv:0804.2541v1 [math-ph]}.

\bibitem{Perlov} 
Perlov L., \emph{Uncertainty principle in loop quantum cosmology by Moyal formalism}, J.~Math.~Phys. {\bf 59} 032304 (2018), \texttt{arXiv:1610.06532v4 [gr-qc]}.

\bibitem{Polymer}
Berra-Montiel J. and Molgado A., \emph{Polymer Quantum Mechanics as a Deformation
Quantization}, Class.~Quantum Grav. {\bf 36} 025001 (2019), \texttt{arXiv:1805.05943v2 [gr-qc]}.

\bibitem{Quasi}
Berra-Montiel J. and Molgado A.,  \emph{Quasi-probability distributions in Loop Quantum
Cosmology}, Class.~Quantum Grav. {\bf 37} 215003 (2020), \texttt{arXiv:2007.01324 [gr-qc]}.

\bibitem{ThiemannDQ1}
Stottmeister A. and Thiemann T., \emph{Coherent states, quantum gravity, and the Born- Oppenheimer approximation. II. Compact Lie groups}, J.~Math.~Phys. {\bf 57} 073501 (2016), \texttt{arXiv:1504.02170 [math-ph]}.

\bibitem{ThiemannDQ2}
Stottmeister A. and Thiemann T., \emph{Coherent states, quantum gravity, and the Born-Oppenheimer approximation. III.: Applications to loop quantum gravity}, J.~Math.~Phys. {\bf 57} 083509 (2016), \texttt{arXiv:1504.02171 [math-ph]}.

\bibitem{EnergyBB}Malkiewicz P. and Piechocki W., \emph{Energy scale of the big bounce}, Phys.~Rev.~{\bf D80} 063506 (2009), \texttt{arXiv:0903.4352 [gr-qc]}.

\bibitem{TurningBB}Dzierzak P., Malkiewicz P. and Piechocki W., \emph{Turning big bang into big bounce: 1. Classical dynamics}, Phys.~Rev.~{\bf D80} 104001 (2009), \texttt{arXiv:0907.3436 [gr-qc]}.

\bibitem{Algebraic}Giesel K. and Thiemann T., \emph{Algebraic quantum gravity (AQG): IV. Reduced phase space quantisation of loop quantum gravity}, Class.~Quantum~Grav.~{\bf 27} 175009 (2010), \texttt{arXiv:0711.0119 [gr-qc]}.




\bibitem{Thiemann}Thiemann T., \emph{Introduction to Modern Canonical Quantum General Relativity} (Cambridge University Press, 2007).

\bibitem{Perez}Perez A., \emph{Regularization ambiguities in loop quantum gravity}, Phys.~Rev.~{\bf D73} 044007 (2006), \texttt{arXiv:gr-qc/0509118}.


\bibitem{Length}Dzierzak P., Jezierski J., Malkiewicz P. and Piechocki W., \emph{The minimum length problem of loop quantum cosmology}, Acta Phys.~Pol.~{\bf B41} 717 (2010), \texttt{arXiv:0810.3172 [gr-qc]}.



\bibitem{Bianchi}Malkiewicz P., Piechocki W. and Dzierzak P., \emph{Bianchi I model in terms of nonstandard loop quantum cosmology: Quantum dynamics}, Class.~Quantum~Grav.~{\bf 28} 085020 (2010), \texttt{arXiv:1010.2930 [gr-qc]}.



\bibitem{TBB2}Malkiewicz P. and Piechocki W., \emph{Turning big bang into big bounce: II. Quantum dynamics}, Class.~Quantum~Grav.~{\bf 27} 225018 (2010), \texttt{arXiv:0908.4029 [gr-qc]}.

\bibitem{Dirac} Dirac P.~A.~M., \emph{Generalized Hamiltonian Dynamics}, Can.~J.~Math.~{\bf 2} 129 (1950).

\bibitem{QSBU}Gazeau J.~P., Mielczarek J. and Piechocki W., \emph{Quantum states of the bouncing universe}, Phys.~Rev.~{\bf D87} 123508 (2013), \texttt{arXiv:1303.1687 [gr-qc]}.

\bibitem{Reed}Reed M. and Simon B., \emph{Methods of Modern Mathematical Physics}, Vol I (Academic Press, United States, 1975).

\bibitem{Evolution} Mielczarek J. and Piechocki W., \emph{Evolution in bouncing quantum cosmology}, Class.~Quantum~Grav.~{\bf 29} 065022 (2012), \texttt{arXiv:1107.4686 [gr-qc]}.

\bibitem{Riemann}Berra-Montiel J. and Molgado A., \emph{Polymeric quantum mechanics and the zeros of the Riemann zeta function}, IJGMMP~{\bf 15} 1850095 (2018), \texttt{arXiv:1610.01957 [math-ph]}.

\bibitem{Dias} Dias N.~C.~ and Prata J.~N., \emph{Wigner functions with boundaries }, J.~Math.~Phys.~{\bf 43} 4602 (2002), \texttt{arXiv:quant-ph/0012140}.

\bibitem{Prata}Dias N.~C.~ and Prata J.~N., \emph{Deformation quantization of confined systems}, Int.~J.~Quantum~Inf.~{\bf 5} 257 (2007), \texttt{arXiv:quant-ph/0612022}.

\bibitem{Weyl}Weyl H., \emph{The Theory of Groups and Quantum Mechanics} (Dover Publications, New York, 1950).

\bibitem{Moyal}J.~E.~Moyal, \emph{Quantum mechanics as a statistical theory}, Proc.~Camb.~Phil.~Soc.~{\bf 45} 99–124 (1949).

\bibitem{Reed II}Reed M. and Simon B., \emph{Methods of Modern Mathematical Physics}, Vol.~II (Academic Press, United States, 1975).

\bibitem{Stratonovich}Stratonovich R.~L., \emph{On the statistical interpretation of quantum theory}, Sov.~Phys.~JETP~{\bf 31} 1012 (1956).

\bibitem{Folland} Folland G.~B.,~\emph{Harmonic Analysis in Phase Space} (Annals of Mathematical Studies Vol. 122) (Princeton University Press, Princeton, 1989).

\bibitem{Curtright}Curtright T.~L., Fairlie D.~B. and Zachos C.~K., \emph{A Concise Treatise on Quantum Mechanics in Phase Space} (World Scientific, Singapore, 2014).

\bibitem{Negativity}Kenfack A. and Zyczkowski K., \emph{Negativity of the Wigner function as an indicator of non-classicality}, J.~Opt.~B: Quantum Semiclass.~Opt.~{\bf 6} 396 (2004), \texttt{arXiv:quant-ph/0406015}.

\bibitem{Gaussian}Mielczarek J.  and  Piechocki W., \emph{Gaussian  state  for  the  bouncing  quantum  cosmology}, Phys.~Rev.~{\bf D86} 8 (2012), \texttt{arXiv:1108.0005 [gr-qc]}.

\bibitem{Numerical}Diener P., Gupt B., Megevand M. and Singh P., \emph{Numerical evolution of squeezed and non-Gaussian states in loop quantum cosmology}, Class.~Quantum~Grav.~{\bf 31} 16 (2014), \texttt{arXiv:1406.1486 [gr-qc]}.

\bibitem{Decoherence}Zurek W.~H., Habib S. and Paz J.~P., \emph{Coherent states via decoherence}, Phys.~Rev.~Lett.~{\bf 70} 1187 (1993).

\bibitem{Hudson}Hudson R.~L., \emph{When is the Wigner quasi-probability density non-negative?}, Rep.~Math.~Phys.~{\bf 6} 249 (1974).

\bibitem{Kiefer}Kiefer C. and Schell C., \emph{Interpretation of the triad orientations in loop quantum cosmology}, Class.~Quantum~Grav.~{\bf 30} 035008 (2013), \texttt{arXiv:1210.0418 [gr-qc]}.






\bibitem{Ariano}D'Ariano G.~M., Mancini S., Man'ko V.~I. and Tombesi P., \emph{Reconstructing the density
operator by using generalized field quadratures}, Quantum~Semiclass.~Opt.~{\bf 8} 1017 (1996), \texttt{ 	arXiv:quant-ph/9606034}.

\bibitem{Homodyne}Vogel K. and Risken H., \emph{Determination of quasiprobability distributions in terms of probability distributions for the rotated quadrature phase}, Phys.~Rev.~{\bf A40} 2847 (1989).

\bibitem{Manko5}Man'ko O.~V., Man'ko V.~I. and Pilyavets O.~V., \emph{Probability Representation of Classical States}, J.~Russ.~Laser~Res.~{\bf 26} 429 (2005).

\bibitem{Exp1}D'Auria V., Fornaro S., Porzio A., Solimeno S., Olivares S., and Paris M.~G.~A., \emph{Full Characterization of Gaussian Bipartite Entangled States by a Single Homodyne Detector}, Phys.~Rev.~Lett.~{\bf 102} 020502 (2009), \texttt{arXiv:0805.1993 [quant-ph]}.

\bibitem{Exp2}Smithey D.~T., Beck M., Raymer M.~G., and Faridani A., \emph{Measurement of the Wigner distribution and the density matrix of a light mode using optical homodyne tomography: Application to squeezed states and the vacuum}, Phys.~Rev.~Lett.~{\bf 70} 1244 (1993). 

\bibitem{Tsymbol}Man'ko O.~V., Man'ko V.~I., Marmo G. and Vitale P., \emph{Star products, duality and double Lie algebras}, Phys.~Lett.~{\bf A360} 522 (2007), \texttt{arXiv:quant-ph/0609041}.



\end{thebibliography}

\end{document}